\documentclass[reprint,amsmath,amssymb,aps,prl,superscriptaddress,]{revtex4-2}

\usepackage{subfiles}
\usepackage{graphicx}% Include figure files
\usepackage{dcolumn}% Align table columns on decimal point
\usepackage{bm}% bold math
\usepackage{color}
\usepackage{braket}
\usepackage{bibunits}
\usepackage{hyperref}
\hypersetup{hidelinks}

\begin{document}
	
\begin{bibunit}
	%\preprint{APS/123-QED}
	
	\title{Large regenerative parametric amplification on chip at ultra-low pump powers}
	
	\author{Yun Zhao}
	\affiliation{Department of Applied Physics and Applied Mathematics, Columbia University, New York, NY 10027, USA}	
	\affiliation{Department of Electrical Engineering, Columbia University, New York, NY 10027, USA}
	\author{Jae K. Jang}
	\affiliation{Department of Applied Physics and Applied Mathematics, Columbia University, New York, NY 10027, USA}
	\author{Xingchen Ji}
	\affiliation{Department of Electrical Engineering, Columbia University, New York, NY 10027, USA}
	\author{Yoshitomo Okawachi}
	\affiliation{Department of Applied Physics and Applied Mathematics, Columbia University, New York, NY 10027, USA}
	\author{Michal Lipson}
	\affiliation{Department of Applied Physics and Applied Mathematics, Columbia University, New York, NY 10027, USA}
	\affiliation{Department of Electrical Engineering, Columbia University, New York, NY 10027, USA}
	\author{Alexander L. Gaeta}
	\email{a.gaeta@columbia.edu}
	\affiliation{Department of Applied Physics and Applied Mathematics, Columbia University, New York, NY 10027, USA}
	\affiliation{Department of Electrical Engineering, Columbia University, New York, NY 10027, USA}

	\date{\today}% It is always \today, today,
	
\begin{abstract}
	Chip-based optical amplifiers can significantly expand the functionalities of photonic devices. In particular, optical-parametric amplifiers (OPAs), with engineerable gain-spectra, are well-suited for nonlinear-photonic applications. Chip-based OPAs typically require long waveguides that occupy a large footprint, and high pump powers that cannot be easily produced with chip-scale lasers. We theoretically and experimentally demonstrate a microresonator-assisted regenerative OPA that benefits from the large nonlinearity enhancement of microresonators and yields a high gain in a small footprint. We achieve 30-dB parametric gain with only 9 mW of cw-pump power and show that the gain spectrum can be engineered to cover telecom channels inaccessible with Er-based amplifiers. We further demonstrate the amplification of Kerr-soliton comb lines and the preservation of their phase properties. Additionally, we demonstrate amplification by injection locking of optical-parametric oscillators, which corresponds to a regenerative amplifier pumped above the oscillation threshold. Novel dispersion engineering techniques such as coupled cavities and higher-order-dispersion phase matching can further extend the tunability and spectral coverage of our amplification schemes. The combination of high gain, small footprint, low pump power, and flexible gain-spectra engineering of our regenerative OPA is ideal for amplifying signals from the nanowatt to microwatt regimes for portable or space-based devices where ultralow electrical power levels are required and can lead to important applications in on-chip optical- and microwave-frequency synthesis and precise timekeeping. 
\end{abstract}

\maketitle

%\tableofcontents
Optical amplifiers are essential components in many optical systems and applications. Amplification is critical in many optical-frequency-comb-related precision metrology tasks, such as $f$-2$f$ locking \cite{Telle_APB_1999}, optical-frequency synthesis \cite{Spencer_Nature_2018}, and microwave generation \cite{Papp_Optica_2014,Lucas_NatComm_2020,Tetsumoto_NatPhot_2021}, for achieving the desired accuracy. Furthermore, gas sensing and optical communications allow limited power in the transmitter due to safety and cost concerns, leading to low signal-to-noise ratio (SNR) due to detector noises, and low-noise optical amplifiers are often needed before the receiver to improve the SNR \cite{Hodgkinson_Misc_2012}. Typically, optical gain is produced via electronic population inversion, which depends critically on material properties. Most gain materials have largely fixed gain spectra and limited  bandwidths. Modification of the gain spectra, when possible, requires significant engineering efforts and results in high production costs. On-chip implementation of optical amplifiers also faces challenges of compatibility issues between different materials. Notable demonstrations of on-chip gain materials include III-V semiconductors integrated on III-V chips \cite{Verma_SPIE_2003,Wan_Misc_2021} or bonded to silicon chips \cite{Roelkens_LaserPhotRev_2010} and Er doping in silicon dioxide, silicon nitride, or lithium niobate \cite{Bradley_LaserPhotRev_2011}. However, to date, these approaches only cover limited spectral ranges and have not been adopted by standard silicon-based foundries. 

Alternatively, optical gain can be produced via nonlinear parametric processes, and can cover a broad spectral range. While optical parametric amplifiers (OPAs) have been routinely used for amplifying signals where conventional material gains are unavailable, due to the limited nonlinearities of most materials, OPAs require long waveguides \cite{Ye_SciAdv_2021,Riemensberger_arXiv_2021} or synchronous-ultrashort-pulse pumping \cite{Foster_Nature_2006,Liu_NatPhot_2010,Wang_JOpt_2015,Ooi_NatComm_2017,Ledezma_Optica_2022}. Optical cavities and resonators have been shown to enhance nonlinearities, with applications including broad-band Kerr-comb generations \cite{Kippenberg_Science_2018,Pasquazi_PhysRep_2018,Gaeta_NatPhot_2019}, low-power frequency conversions \cite{Li_NatPhot_2016,Chen_Optica_2019,Lu_Optica_2019}, and photon-level nonlinear interactions \cite{Lu_Optica_2020}. Multipass pseudo-cavities \cite{Backus_RSI_1998} and singly resonant cavities \cite{Ilday_OL_2006} have been used to develop bulk pulse-pumped OPA systems. However, such approaches still require high peak powers and are not easily adaptable to chip-based systems. Since typical chip-scale single-mode lasers have output powers in the milliwatt regime \cite{Han_AOP_2022}, a high-gain low-pump-power OPA is desirable for numerous applications requiring low SWaP (size, weight, and power).

\begin{figure*}
	\centering
	\includegraphics{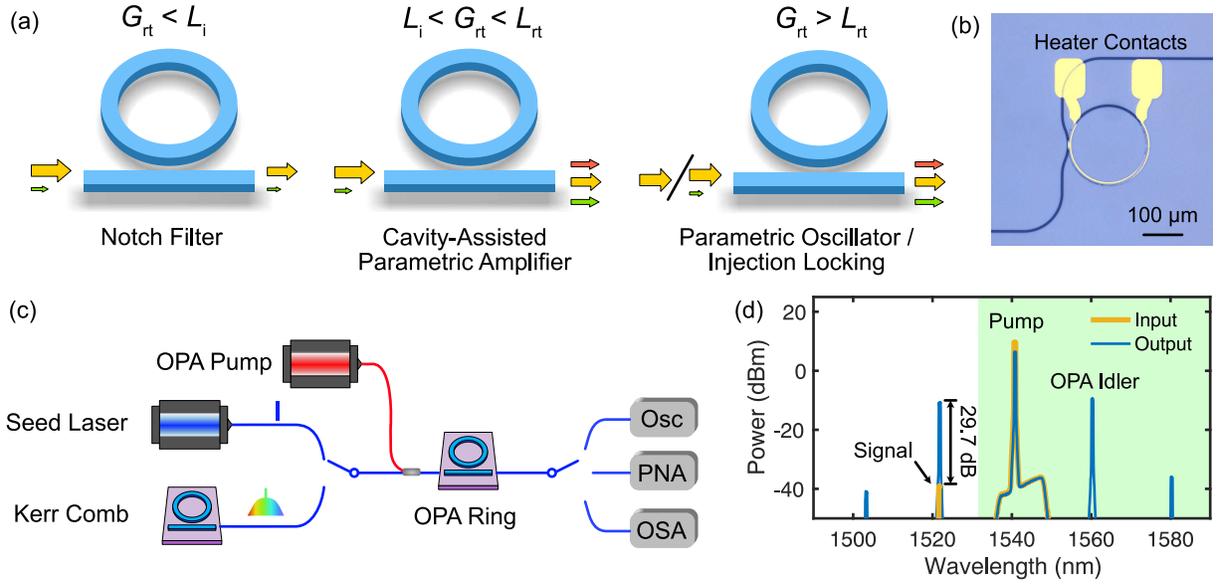}
	\caption{Schematic of parametric amplification with microresonators. (a), Different operation regimes of cw-pumped microresonators, where $G_\text{rt}$ is the roundtrip parametric gain, $L_\text{i}$ is the intrinsic roundtrip loss, and $L_\text{rt}$ is the total roundtrip loss. The microresonator functions as a notch filter, parametric amplifier, and parametric oscillator for $G_\text{rt} < L_\text{i}$, $L_\text{i}<G_\text{rt} < L_\text{rt}$, and $G_\text{rt} > L_\text{rt}$, respectively. The parametric oscillator can also be injection-locked by an external signal, which alters the phase properties of the output while keeping the power level nearly identical. (b), Optical microscope image of the experimental device. The footprint can be further reduced by replacing the ring resonator with a spiral resonator. (c), Schematic of the experiment. We use our OPA to amplify a seed laser or lines of a Kerr comb. The output is analyzed with an oscilloscope (Osc), a phase-noise-analysis (PNA) system, and an optical spectrum analyzer (OSA). (d), Experimental input and output spectra of the OPA system. The green-shaded region corresponds to the spectral range covered by Er-doped fiber amplifiers.}
	\label{figScheme}
\end{figure*}

We propose two microresonator-assisted light amplification schemes on a chip that are readily attainable with standard microresonators and allow for high gain at low pump powers. It is well-known that when the round-trip gain $G_\mathrm{rt}$ exceeds the round-trip loss $L_\mathrm{rt}$ inside a cavity, the net gain results in optical oscillation, instead of amplification. However, we show that even without a net gain inside the cavity the field in the bus waveguide can be amplified, which only requires the round-trip gain being larger than the \textit{intrinsic} loss $L_\mathrm{i}$ of the cavity. In this scheme, the signal effectively experiences multi-pass gain, which far exceeds that in a singly-resonant configuration, allowing for high gain coefficients. This OPA scheme is similar to the regenerative amplification regime of laser cavities \cite{Siegman_Lasers} and microwave parametric amplification of nonlinear LC oscillators \cite{Aumentado_Misc_2020}. Alternatively, regenerative amplifiers can operate with the pump power above the oscillation threshold, which is also referred to as injection locking \cite{Siegman_Lasers}. In this regime, the output does not vanish when the input is off, thus erasing amplitude-encoded information. However, we show that the output phase characteristics of the injection-locked output faithfully follow that of the input signal. Both regenerative OPA and injection locking can be readily realized in nonlinear microresonators with a pump power at sub-milliwatt levels in state-of-the-art high-$Q$ devices \cite{Kippenberg_PRL_2004,Ji_Optica_2017,LuX_Optica_2019}. The gain spectra of these approaches can be readily tuned via dispersion engineering without changing the fabrication process, allowing for amplification in regimes where material gain is not accessible. The regenerative OPA configuration has previously been used for frequency conversion, with output signal powers being much less than the input signal power \cite{Turner_OE_2008,Pasquazi_OE_2010,Li_OE_2011,Lu_NatPhot_2019} due to the loss in the resonators. Recently, microresonator-assisted parametric amplification was experimentally observed with 1.4-dB gain at 5-mW pump power \cite{Zheng_Light_2016}, 20-dB gain at 300-mW pump power \cite{Geng_OL_2019}, and 5-dB gain at 19-mW pump power \cite{Black_arXiv_2021}, respectively. However, the origin of the gain was inaccurately attributed to the round-trip parametric gain being higher than the round-trip loss \cite{Geng_OL_2019}, which corresponds to the transient regime of an OPO and thus cannot be applied to the suitable design of an amplifier. This misunderstanding can potentially mislead future OPA designs. Here, we theoretically clarify the origin of parametric gain and use it as a guide to experimentally demonstrate up to 30-dB gain with 9 mW of pump power in a silicon-nitride microresonator, which represents a 10$\times$ higher gain at 30$\times$ lower pump power compared to the previous work. Importantly, our demonstration represents the first OPA design that is compatible with typical edge-coupled or heterogeneously integrated chip-scale semiconductor lasers, establishing microresonator-assisted regenerative OPA as a highly practical device for on-chip integration.

\begin{figure*}
	\centering
	\includegraphics{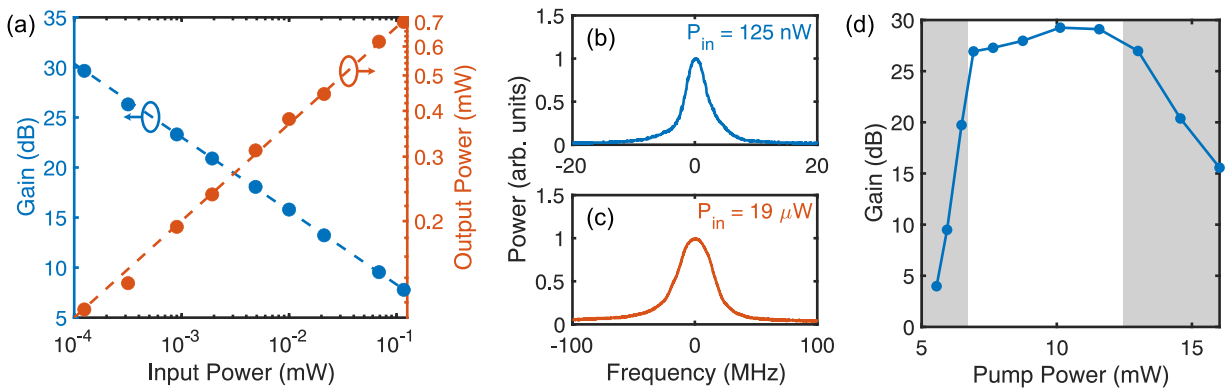}
	\caption{Characterization of the microresonator-assisted regenerative OPA gain. (a), Experimental gain and saturated output power of the OPA system, where dashed lines are fits to an exponential function. (b), (c) the gain spectral response for 125 nW input power and 19 $\mu$W input power, respectively. (d), The OPA gain coefficient as a function of pump power with an input-signal power of 166 nW. The shaded region on the left corresponds to pump power below the OPO threshold. The central region corresponds to pump power being higher than the threshold power. However, the detuning is chosen such that the cavity does not oscillate. The shaded region on the right corresponds to the first OPO sideband being different from the signal wavelength being amplified due to increased pump power. }
	\label{figGainCharact}
\end{figure*}

Our regenerative OPA scheme is based on below-threshold degenerately-pumped four-wave mixing (DFWM) in a microresonator, and the same principle can be applied to all parametric gain processes. For simplicity, we represent the nonlinear interaction of the signal mode as a roundtrip power gain term $G_\mathrm{rt}$ and a detuning term $\Delta_\mathrm{NL}$ (see Supplementary Material). The small-signal power gain for the output field can be expressed as,
%\begin{align}
%	G = \frac{(L_\mathrm{rt}+G_\mathrm{rt}-2L_\mathrm{i})^2\mathcal{F}^2+4(\Delta-2\Gamma|A|^2-\Delta_\mathrm{NL})^2}{(L_\mathrm{rt}-G_\mathrm{rt})^2\mathcal{F}^2+4(\Delta-2\Gamma|A|^2-\Delta_\mathrm{NL})^2},\label{eq_gain}
%\end{align}
\begin{align}
		G = 1+\frac{4(L_\mathrm{rt}-L_\mathrm{i})(G_\mathrm{rt}-L_\mathrm{i})\mathcal{F}^2}{(L_\mathrm{rt}-G_\mathrm{rt})^2\mathcal{F}^2+4(\Delta-2\Gamma|A|^2-\Delta_\mathrm{NL})^2},\label{eq_gain}
\end{align}
where $L_\mathrm{i}$ is the intrinsic roundtrip loss, $L_\mathrm{rt}$ is total roundtrip loss, $\Gamma$ is the cavity-enhanced nonlinear coefficient, $\Delta$ is the signal detuning, $\mathcal{F}$ is the free-spectral range (FSR), and $A$ is the intracavity pump amplitude such that $|A|^2$ is normalized to power. Inspection of Eq. (\ref{eq_gain}) shows that the system has a net gain $G>1$ if $G_\mathrm{rt} > L_\mathrm{i}$, that is, the nonlinear gain is larger than the intrinsic loss. The second terms in the numerator and denominator correspond to the detuning from the cavity resonance, which impact the maximum gain and bandwidth of the gain profile. As $G_\mathrm{rt}$ approaches $L_\mathrm{rt}$, the cavity has an effective ultra-high $Q$ exceeding that of a lossless cavity with the same bus-cavity coupling rate. The gain process can be attributed to the extraction of the strong cavity-enhanced field at a high rate. An ideal phase-insensitive optical amplifier has a noise figure NF = 2 in the high-gain regime, where the NF is defined as the ratio between the SNRs of the output and input signals, and SNR is defined as the ratio between the square of the optical power and the variance of the optical power (see Supplementary Material). In addition, the NF of 2 can be realized only when the bandwidth of the noise is filtered to match that of the signal, which is then limited by the quantum noise \cite{Haus_Noise}. Similar to many cavity-based quantum systems, the quantum-noise-limited behavior of the regenerative OPA depends on the coupling condition of the cavity \cite{Collett_PRA_1984,Vernon_PRA_2015,Zhao_PRL_2020}. As detailed in Supplementary Material, NF of 2 can be achieved in our OPA configuration when the cavity is strongly over coupled, that is, $L_\mathrm{rt}\gg L_\mathrm{i}$. For a critically coupled cavity ($L_\mathrm{rt}=2L_\mathrm{i}$), which minimizes the input pump power, the lowest achievable NF is 4, which is due to the additional fluctuations induced by the scattering losses. In practical systems, the NF is also affected by the pump intensity and phase noise, and the thermal noise of the microresonator, which can be mitigated by using lower-$Q$ cavities and higher pump powers.

\begin{figure*}
	\centering
	\includegraphics{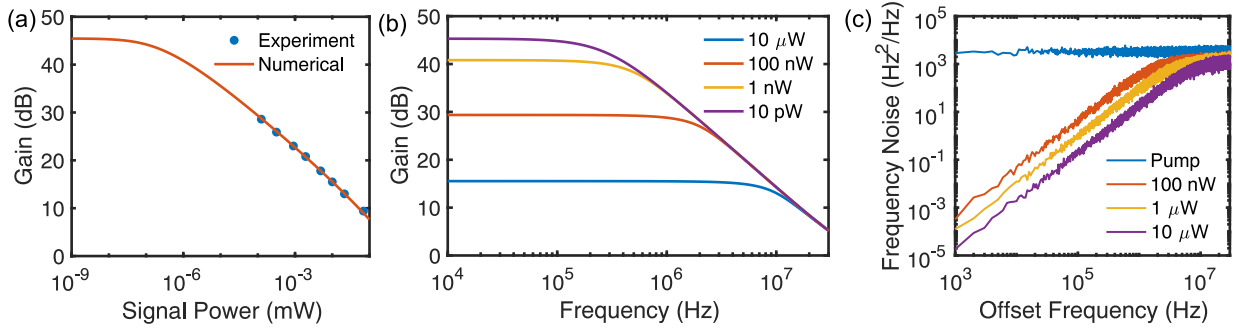}
	\caption{Simulations of the microresonator-assisted regenerative OPA system. (a), Simulated gain with close-to-threshold pump power, showing a small-signal gain of 45 dB and an onset of gain saturation at 100 pW. The experimental data from Fig. \ref{figGainCharact}(a) is overlaid on the figure, where the pump detuning (29 MHz) is used as the only free parameter to fit the data. (b), Simulated gain bandwidth for varying seed powers. All input levels have a constant GBW of 1.9 GHz. (c), Simulated frequency-noise transfer, showing large pump-noise isolation at lower offset frequencies, which diminishes at higher offset frequencies.}
	\label{figTheory}
\end{figure*}

In our experimental demonstration, we design our regenerative OPA to operate near 1522 nm, which is outside the Er-gain window. The microresonator cross-section is 730 $\times$ 1500 nm with a radius of 100 $\mu$m (FSR = 220 GHz) and has a loaded and intrinsic $Q$ of $1.4\times10^6$ and $3.6\times10^6$, respectively. We use a pump at 1540.8 nm, corresponding to a peak gain at 1521.8 nm and 1560.3 nm. As shown in Fig. \ref{figScheme}(d), the orange traces correspond to the input pump and signal fields, and the blue traces correspond to the output fields in the OPA regime. With 9 mW of pump power, the 125-nW input signal is amplified by a factor of $\approx$ 1000. An almost equally strong idler field is generated due to energy conservation of the DFWM process. The asymmetry of signal and idler power is due to different bus-ring coupling efficiency and output collection efficiency in the experiment. We further observe cascaded lines due to the high effective nonlinearity of our system. The cascading can be suppressed by creating strong dispersive effects such as local mode coupling \cite{Heuck_NJP_2019} or phase matching with different mode families \cite{Perez_NatComm_2023}. Since high gain requires the system to exhibit high $Q$, there is a trade-off between the gain and the bandwidth, which is typical for many types of amplifiers. Figures \ref{figGainCharact}(b) and \ref{figGainCharact}(c) show the frequency response of the 125 nW and 19 $\mu$W input signals with a gain coefficient of 30 dB and 13 dB, respectively. The gain-bandwidth product (GBW) can be further improved by increasing the bus-ring coupling coefficient and pump power. Due to the high gain of the regenerative OPA, it operates in the saturation regime even for a 125-nW input signal, which is the current limit of our photodetector sensitivity. We measure the output powers for varying input powers and find that $P_o \propto P_i^{0.26}$ where $P_o$ and $P_i$ are the output and input powers, respectively [Fig. \ref{figGainCharact}(a)]. Figure \ref{figGainCharact}(d) shows the gain as a function of pump power with an input-signal power of 166 nW. The gain quickly rises as the pump power approaches the OPO threshold and increases slowly after the pump power crosses the threshold. The system does not oscillate due to the carefully chosen pump-cavity detuning. At high pump powers, the first OPO sideband shifts farther from the pump, leading to a reduced gain at the original OPA resonance. We also numerically investigate the gain for a larger range of input powers as shown in Fig. \ref{figTheory}(a) (see Supplementary Material). We observe a small-signal-gain of 45 dB and the onset of gain saturation at 100 pW [Fig. \ref{figTheory}(a)]. The gain reduces for higher seed powers due to saturation. However, the GBW remains constant at 1.9 GHz [Fig. \ref{figTheory}(b)]. 

\begin{figure*}
	\centering
	\includegraphics{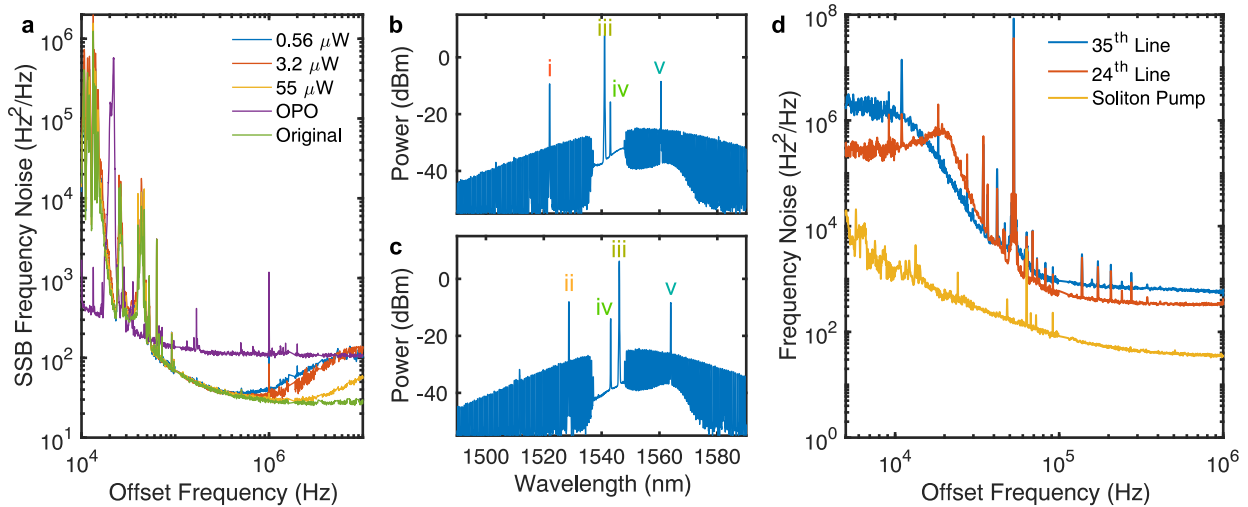}
	\caption{Phase noise characterization of the regenerative OPA system. (a), Noise of the amplified signal for different input-signal levels. The OPO noise, which also largely agrees with the pump noise, is shown as a comparison. The high-frequency deviations from the original phase noise are due to the relatively low optical power in the measurement system. (b), (c), Output spectra of soliton comb lines amplified by the regenerative OPA. (i) and (ii) are the amplified comb lines at 1522.0 and 1528.7 nm, respectively. (iii) is the OPA pump, (iv) is the residual soliton pump, and (v) is the OPA idler. (d), Measured phase noise of the amplified comb lines.}
	\label{figFrequencyNoise}
\end{figure*}

We characterize the phase-noise properties of the regenerative OPA system, which is important for applications including $f-2f$ stabilization \cite{Telle_APB_1999}, optical frequency division \cite{Papp_Optica_2014}, and optical frequency synthesis \cite{Spencer_Nature_2018}. We use a signal laser at 1560.3 nm that is lower noise than the pump laser at 1540.8 nm, and characterize the frequency noise of amplified signals using a delayed self-heterodyne system \cite{Camatel_JLT_2008}. The phase noise of the seed laser is characterized at a power level of 5 mW, and the seed is strongly attenuated before being sent into the OPA. Figure \ref{figFrequencyNoise}(a) shows the phase noise of the amplified signals with varying input power levels. The amplified signals have the same noise characteristics as the original signal at lower offset frequencies. The deviation at the high offset frequencies is largely due to the noise floor of the photodetector. For comparison, we pump the microresonator above the OPO threshold and characterize the corresponding OPO noise. The OPO noise largely follows that of the pump laser (see Supplementary Material), which is generally much higher than the seed laser and has distinctive high-noise sidebands. As shown in Fig. \ref{figFrequencyNoise}(a), the phase noise of the amplified signal is not affected by the pump noise, where the suppression of the noise feature at 22 kHz corresponds to a minimum of 31-dB pump-noise isolation. This pump-noise isolation is also verified with our numerical model. As shown in Fig. \ref{figTheory}(b) (see Supplementary Material), larger pump-noise suppression is achieved at lower offset frequencies and higher seed powers.

Kerr combs allow for the on-chip generation of broadband modelocked combs with a cw pump laser and have shown great promise for frequency-synthesis applications \cite{Kippenberg_Science_2018,Gaeta_NatPhot_2019}. However, in the soliton-modelocking regime, most Kerr combs have low comb line powers, which could require off-chip amplification for many applications \cite{Papp_Optica_2014,Spencer_Nature_2018,Lucas_NatComm_2020,Tetsumoto_NatPhot_2021}. We experimentally show that our regenerative OPA system can achieve strong on-chip amplification at specified soliton comb lines. We generate the Kerr soliton using a microresonator with a cross-section of 730 $\times$1700 nm and a FSR of 77 GHz. Figures \ref{figFrequencyNoise}(b) and \ref{figFrequencyNoise}(c) show the sech$^2$ soliton spectral shape where the center part is filtered due to the OPA-pump combination using a wavelength-division multiplexer (WDM). The WDM can also be replaced by an on-chip add-drop-ring filter that does not affect any comb lines. We use an OPA pump power of 8.6 mW and a soliton pump power of 340 mW. We jointly tune the OPA pump wavelength and the OPA ring resonance (via an on-chip heater) to achieve amplification of different comb lines. As shown in Figs. \ref{figFrequencyNoise}(b) and \ref{figFrequencyNoise}(c), traces (i) and (ii) correspond to amplified comb lines of the 35$^\text{th}$ (1522.0 nm) and 24$^\text{th}$ modes (1528.7 nm), with a gain factor of 22 dB and 21 dB, respectively. The corresponding idler modes are shown as trace (v), which do not overlap with any comb modes. The amplification allows us to characterize the phase noise of the high-order comb lines, which is not detectable with our measurement system at their original power levels. As shown in Fig. \ref{figFrequencyNoise}(d), the comb-line noise increases with the mode order, which is due to the cascading of FSR fluctuations. Consequently, the FSR fluctuations can be significantly reduced if high-order comb lines are measured and stabilized, which is the critical part of frequency synthesis. Our OPA provides an approach for high SNR measurement of the noise at the low-power comb lines, which is otherwise strongly limited by detector noise.

\begin{figure}
	\centering
	\includegraphics{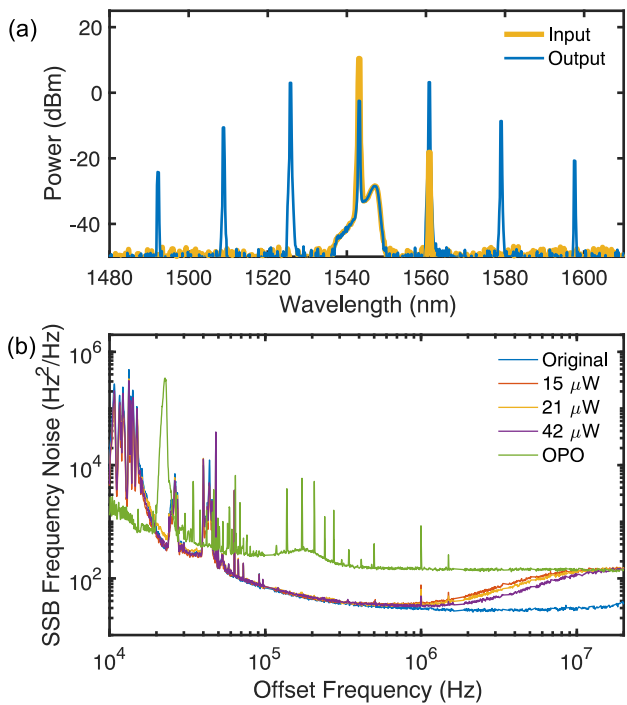}
	\caption{Optical amplification via injection locking. a, The optical spectra of the off-resonance pump and signal (orange), and the injection-locked OPO (blue), showing a pump-to-signal conversion efficiency of -7 dB. b, Frequency noise of the injection-locked OPO versus free-running OPO. Different injection laser strengths are compared.}
	\label{figInjection}
\end{figure}

For applications only requiring phase information, optical amplification can also be performed via injection locking, which is conceptually closely related to regenerative amplification \cite{Siegman_Lasers}. In such a process, A weak laser within a certain frequency range of a free-running optical tone is injected onto the cavity, where the free-running tone is subsequently phase-locked to the injecting laser. The capturing range (allowable frequency separation between the seed and the free-running tone) depends on the strength of the seed laser and the cavity-coupling condition. This procedure has been demonstrated for laser cavities, with applications in phase-encoded telecommunications \cite{Kobayashi_IEEE_1982} and interferometric gravitational-wave sensors \cite{Kwee_OE_2012}. Here, we show that it can also be realized in the OPO regime, where the free-running tone is the signal of the OPO. The experiment setup of injection locking is identical to that of the regenerative OPA [Fig. \ref{figScheme}(c)]. However, in the OPA regime, the pump is tuned below the oscillation threshold. In contrast, in the injection-locking regime, the OPO is generated before we send in the seed laser. To better demonstrate the frequency noise characteristics, we use a probe laser at 1560 nm, which has lower noise than the pump laser. However, dispersion engineering can also be applied to amplify fields at other frequencies. With a pump power of 11 mW, we generate an OPO with 2.1 mW of signal/idler power, corresponding to a -7-dB power-conversion efficiency for each wavelength. The free-running phase noise is shown as the green trace in Fig. \ref{figInjection}(b), which largely follows the pump noise (see Supplementary Material). We vary the injection power at 1560.8 nm and observe the change of phase noise. Experimentally, a minimum seed power of 15 $\mu$W is required to lock the OPO signal to the seed laser, which corresponds to a maximum power gain of 21 dB [Fig. \ref{figInjection}(a)]. As shown by the red trace in Fig. \ref{figInjection}(b), the injection forces the phase noise of the OPO signal to follow that of the seed laser at offset frequencies below 300 kHz, where the original noise features are completely removed. Notably, the removal of the free-running-OPO noise peak at 23 kHz corresponds to a $>$ 30 dB pump-noise isolation. At higher offset frequencies, the noise of the locked signal approaches that of the free-running OPO, which can be mitigated with higher seed-laser powers [Fig. \ref{figInjection}(b)].

Our system could be used in applications requiring high amplification in an integrated photonic platform in which limited electrical power is available (e.g., portable, space-based systems). Our system can, in principle, operate at the quantum-amplifier-noise limit and could be used to amplify signals from the nanowatt to the microwatt regime. The low saturation power can be partially addressed by the injection locking scheme. However, the tradeoff of this high gain at low powers is the narrow gain bandwidth, which would make it unsuitable for applications related to high-speed pulses or information encoded in multiple amplitude levels. Nonetheless, for many applications this tradeoff is immaterial, such as for on-chip time and frequency metrology. For example (see Supplementary Material for more applications), one promising route towards on-chip optical atomic clocks requires a fully stabilized Kerr comb and a sufficiently strong comb line to probe an on-chip atomic cell \cite{Papp_Optica_2014,Yu_PRAppl_2019}. Comb stabilization requires the frequency doubling of a low-frequency comb line to overlap with a high-frequency comb line, which are necessarily an octave apart. Due to the sech$^2$ profile of Kerr solitons, the two comb lines typically have sub-microwatt level powers, even with dispersive-wave formation \cite{Pfeiffer_Optica_2017}. Efficient frequency doubling can be achieved with 100-$\mu$W level power in state-of-the-art periodically-poled-lithium-niobate cavities \cite{Lu_Optica_2019}, which can be reached by sub-microwatt-level comb lines with our regenerative OPA. Alternatively, frequency doubling can be achieved with the photogalvanic effect \cite{Anderson_OL_1991} in the same SiN platform for Kerr comb generation, albeit at a lower efficiency. Due to the absence of inhomogeneously-broadening-induced material fluorescence, and the low added quantum noise, our OPA scheme is uniquely suited for amplifying the low-level second-harmonic signal due to inefficient frequency doubling, which can be at nanowatt level.

In conclusion, we have theoretically and experimentally demonstrated two approaches to achieve high on-chip optical-parametric amplification, namely regenerative OPA and injection-locked OPO, which allow for spectral coverage not possible with active gain materials. We have shown for a regenerative OPA that a high gain can be achieved with pump powers near to the oscillation threshold with a tradeoff of gain bandwidth and saturation power. At higher pump powers, the pump-cavity detuning can be used to control the OPA gain. However, the maximum gain does not increase at even higher pump powers. Due to the relatively low gain bandwidth, both the pump and the resonance need to be fine-tuned to access the high-gain regime for the signal, which can be achieved with on-chip heaters and pump-current tuning. Benefiting from cavity enhancement, our scheme allows for higher gain at much lower pump powers and smaller footprints than previous on-chip OPA demonstrations \cite{Foster_Nature_2006,Liu_NatPhot_2010,Zheng_Light_2016,Geng_OL_2019,Ye_SciAdv_2021,Riemensberger_arXiv_2021}, and the footprint of our device is also significantly smaller than most devices with comparable material gain \cite{Liu_Science_2022}. Particularly, the pump-power level for our regenerative-OPA and injection-locked-OPO schemes is fully compatible with typical chip-scale single-mode semiconductor lasers. Our method also applies to other nonlinear gain processes, such as $\chi^{(2)}$ processes. Novel dispersion engineering techniques can be applied to expand the gain region. For example, coupled-ring structures can be used to locally tune the dispersion \cite{Xue_LaserPhotRev_2015,Kim_OL_2019,Okawachi_OL_2022}, and we observe active tuning of the peak gain by $\approx$ 90 nm on each side of the pump (Supplementary Material). Higher-order-dispersion phase matching can allow for peak gain far away from the pump \cite{Harvey_OL_2003,Sayson_NatPhot_2019,LuX_Optica_2019,Domeneguetti_Optica_2021}, which can provide high-gain amplifiers for mid-infrared or ultraviolet regimes (Supplementary Material). We observe that the phase noise of the amplified signal is well-isolated from those of the pump. We have shown that our regenerative OPA scheme is suitable for the amplification of Kerr-comb lines, which have intrinsically low powers, and this technology can help expand the scope of fully integrable Kerr-comb applications, including atomic clocks \cite{Papp_Optica_2014}, optical frequency synthesis \cite{Spencer_Nature_2018}, and ultra-low-noise microwave generation \cite{Lucas_NatComm_2020,Tetsumoto_NatPhot_2021}.
\\[6pt]
\textbf{Acknowledgements.} 
\begingroup
\fontsize{8pt}{6pt}\selectfont
This work was performed in part at the Cornell Nano-Scale Facility, which is a member of the National Nanotechnology Infrastructure Network, supported by the NSF, and at the CUNY Advanced Science Research Center NanoFabrication Facility.
\endgroup
\\[6pt]
\textbf{Funding}. 
\begingroup
\fontsize{8pt}{6pt}\selectfont
This work was supported by Army Research Office (ARO) (Grant No. W911NF-21-1-0286), Air Force Office of Scientific Research (AFOSR) (Grant No. FA9550-20-1-0297), and Defense Advanced Research Projects Agency of the U.S. Department of Defense (Grant No. HR0011-22-2-0007 ). 
\endgroup
\\[6pt]
\textbf{Disclosures.} 
\begingroup
\fontsize{8pt}{6pt}\selectfont
The authors declare no conflicts of interest.
\endgroup
\\[6pt]
\textbf{Data availability.}
\begingroup
\fontsize{8pt}{6pt}\selectfont
Data underlying the results presented in this paper are not publicly available at this time but may be obtained from the authors upon reasonable request.
\endgroup
\\[6pt]
\textbf{Supplemental document.}
\begingroup
\fontsize{8pt}{6pt}\selectfont
See Supplement 1 for supporting content.
\endgroup

\end{bibunit}

	\newpage
	\onecolumngrid
	\setcounter{equation}{0}
	\setcounter{figure}{0}
	\setcounter{table}{0}
	\setcounter{page}{1}
	\setcounter{section}{0}
	\renewcommand{\theequation}{S\arabic{equation}}
	\renewcommand{\thefigure}{S\arabic{figure}}
	\renewcommand{\bibnumfmt}[1]{[S#1]}
	\renewcommand{\citenumfont}[1]{S#1}

\begin{bibunit}
	%\preprint{APS/123-QED}
	\begin{center}
	\textbf{\large Large regenerative parametric amplification on chip at ultra-low pump powers: supplemental document}
	\end{center}
%	\date{\today}% It is always \today, today,
	
%\maketitle

%\tableofcontents
\section{Classical Model of OPA.} 
The classical dynamics of our OPA system are governed by 
\begin{align}
	&\frac{dA}{dt} = -\frac{\alpha}{2}A -i\Delta_AA+i\Gamma(|A|^2+2|B|^2+2|C|^2)A+i2\Gamma A^\ast BC + \sqrt{\alpha_1 \mathcal{F}}A_i,\label{eq_Classical1}\\
	&\frac{dB}{dt} = -\frac{\alpha}{2}B -i\Delta_BB+i\Gamma(2|A|^2+|B|^2+2|C|^2)B+i\Gamma A^2C^\ast +\sqrt{\alpha_1\mathcal{F}}B_i,\label{eq_Classical2}\\
	&\frac{dC}{dt} = -\frac{\alpha}{2}C -i\Delta_CC+i\Gamma(2|A|^2+2|B|^2+|C|^2)C+i\Gamma A^2B^\ast,\label{eq_Classical3}
\end{align}
where $A$, $B$, and $C$ are the intracavity field amplitudes of the pump, signal, and idler fields, respectively, $\alpha$ is the loss rate of the cavity, $\Delta_A$, $\Delta_B$, and $\Delta_C$ are the detunings of the pump, signal, and idler fields, respectively, $\Gamma$ is the cavity-enhanced nonlinear coefficient, $\alpha_1$ is the bus-ring coupling rate, $\mathcal{F}$ is the FSR, and $A_i$ and $B_i$ are the input fields of the pump and signal, respectively. We normalize the fields such that the square norm of the amplitudes corresponds to power. The cavity-enhanced nonlinear coefficient $\Gamma = \gamma L\mathcal{F}$, where $\gamma$ is the nonlinear coefficient, and $L$ is the cavity length. We also note that the idler detuning $\Delta_C$ is fully determined by $\Delta_A$, $\Delta_B$, and the cavity dispersion. 

We first look at the gain of a weak input signal. We limit the amplitude $B_i$ such that $|B|^2, |C|^2 \ll |A|^2$ at steady state. The nonlinear term in Eq. (\ref{eq_Classical2}) can be written as
\begin{align}
	i\Gamma A^2C^\ast =& \frac{\frac{\alpha}{2}\Gamma^2|A|^4}{\frac{\alpha^4}{4}+(\Delta_C-2\Gamma|A|^2)^2}B+ i\frac{(\Delta_C-2\Gamma|A|^2)\Gamma^2|A|^4}{\frac{\alpha^2}{4}+(\Delta_C-2\Gamma|A|^2)^2)}B\notag\\
	=& \frac{G_\mathrm{rt}\mathcal{F}}{2}B + i\Delta_\mathrm{NL} B,\label{eq_Nonlinear}
\end{align}
where $G_\mathrm{rt}$ is the induced roundtrip power gain and $\Delta_\mathrm{NL}$ is the induced detuning. This is due to the real part of the coefficient before $B$ compensating for the effect of $\alpha$, which represents loss, and the imaginary part of the coefficient before $B$ compensating for the effect of $\Delta_B$, which represents detuning. Hence, Eq. (\ref{eq_Nonlinear}) indicates that the nonlinear term $\Gamma A^2C^\ast$ can be effectively viewed as contributing to gain and detuning. The output field can be found as
\begin{align}
	B_o &= B_i - \sqrt{\frac{\alpha_1}{\mathcal{F}}}B \notag\\
	&= \frac{\alpha-2\alpha_1-G_\mathrm{rt}\mathcal{F}+i2(\Delta_B-2\Gamma |A|^2-\Delta_\mathrm{NL})}{\alpha-G_\mathrm{rt}\mathcal{F}+i2(\Delta_B-2\Gamma |A|^2-\Delta_\mathrm{NL})}B_i.
\end{align}
With $L_\mathrm{rt} = \alpha/\mathcal{F}$ and $L_\mathrm{i} = (\alpha-\alpha_1)/\mathcal{F}$, we get the net power gain as shown in Eq. (1) in the main text.

In general, Eqs. (\ref{eq_Classical1}-\ref{eq_Classical3}) can be solved numerically to fully incorporate gain saturation. The input fields can also be treated as stochastic variables to simulate noise properties. We numerically simulate a systems with $\mathcal{F}$ = 200 GHz, $\alpha = 2\pi\times $150 MHz, $\alpha_1 = 2\pi\times 100$ MHz, $\Gamma = 1.38\times 10^8$ W$^{-1}$s$^{-1}$, $\Delta_A = 2\pi\times 28$ MHz. We assume a group-velocity dispersion of -50 ps$^2$/km which corresponds to a detuning relation $\Delta_B+\Delta_C = 2\Delta_A + 1.3\times10^9$ s$^{-1}$. To explore the gain saturation at a pump power of 9 mW, we keep $\Delta_B = \Delta_C$, which corresponds to the peak gain detuning, and we vary the input signal power. The corresponding result is shown in Fig. 3(a). To simulate the phase noise properties, we keep $\Delta_B = \Delta_C$ and let $A_i$ be a random process corresponding to a Lorentzian spectral shape with a full-width-at-half-maximum linewidth of 10 kHz. The corresponding result is shown in Fig. 3(b).

\section{Quantum Model of OPA.} 
For a weak signal input, that is, without gain saturation, the quantum dynamics of the system can be modeled as
\begin{align}
	&\frac{d\hat{b}}{dt} = -\frac{\alpha_1+\alpha_2}{2}\hat{b}-i\Delta_B\hat{b}+i2\Gamma |A|^2\hat{b}+i\Gamma A^2\hat{c}^\dagger+\sqrt{\alpha_1\mathcal{F}\hbar\omega}\hat{b}_i+\sqrt{\alpha_2\mathcal{F}\hbar\omega}\hat{l},\label{eq_Quantum1}\\
	&\frac{d\hat{c}^\dagger}{dt} = -\frac{\alpha_1+\alpha_2}{2}\hat{c}^\dagger+i\Delta_C\hat{c}^\dagger-i2\Gamma |A|^2\hat{c}^\dagger-i\Gamma {A^\ast}^2\hat{b}+ \sqrt{(\alpha_1+\alpha_2)\mathcal{F}\hbar\omega}\hat{g}^\dagger,\label{eq_Quantum2}
\end{align}
where $|A|^2$ is the intracavity pump power, $\hat{b}$ and $\hat{c}$ are the annihilation operators of the cavity modes, $\alpha_1$ is the coupling rate, $\alpha_2$ is the loss rate, $\Delta_B$ and $\Delta_C$ are the detunings of the signal and idler, $\Gamma$ is the cavity-enhanced nonlinear coefficient, $\mathcal{F}$ is the FSR, $\omega$ is the pump frequency, $\hat{b}_i$ is the annihilation operator of the input field, $\hat{l}$ is the annihilation operator of the reservoir modes coupled to $\hat{b}$ via scattering, and $\hat{g}$ is the annihilation operator corresponding to all dissipation mechanisms of mode $\hat{c}$. In our treatment, the intracavity fields serve as intermediate steps to connect the input and output fields via $\hat{b}_o = \hat{b}_i-\sqrt{\alpha_1/(\mathcal{F}\hbar\omega)}\hat{b}$, where $\hat{b}_o$ is the annihilation operator of the output field. We choose the convention such that the intracavity fields are normalized to power, which is consistent with the classical equations. Equations (\ref{eq_Quantum1}) and (\ref{eq_Quantum2}) can be solved in the frequency domain, where we define 
\begin{align}	
	\tilde{o}(\Omega) = \frac{1}{\sqrt{2\pi}}\int \hat{o}(t)e^{i\Omega t}dt,
\end{align}
with $\hat{o}$ being any annihilation operator. For simplicity, we let $\Delta_B=\Delta_C=2\Gamma|A|^2$, which corresponds to the peak gain. With some calculation, we can find that
\begin{align}
	\tilde{b}_o(\Omega) = G_1(\Omega)\tilde{b}_i(\Omega)+G_2(\Omega)\tilde{l}(\Omega)+G_3(\Omega)\tilde{g}^\dagger(-\Omega),
\end{align}
where
\begin{align}	
	&G_1(\Omega) = \frac{\Gamma^2|A|^2+(\frac{\alpha_1+\alpha_2}{2}-i\Omega)(\frac{\alpha_1-\alpha_2}{2}+i\Omega)}{(\frac{\alpha_1+\alpha_2}{2}-i\Omega)^2-\Gamma^2|A|^2},\\
	&G_2(\Omega) = \frac{\sqrt{\alpha_1\alpha_2}(\frac{\alpha_1+\alpha_2}{2}-i\Omega)}{(\frac{\alpha_1+\alpha_2}{2}-i\Omega)^2-\Gamma^2|A|^2},\\
	&G_3(\Omega) = \frac{\Gamma |A| \sqrt{\alpha_1(\alpha_1+\alpha_2)}}{(\frac{\alpha_1+\alpha_2}{2}-i\Omega)^2-\Gamma^2|A|^2}.
\end{align}
We have absorbed extra phase terms into the operators, which does not change their commutation relations. We consider an optical pulse with a spectral envelope $s(\Omega)$, where $\int s^\ast s d\Omega = 1$. For simplicity, we let the bandwidth of $s$ be much smaller than the gain bandwidth. The input-output relation for this pulse mode can be written as
\begin{align}
	\hat{b}_{os} = G_1(0)\hat{b}_{is} + G_2(0)\hat{l}_s + G_3(0)\hat{g}_s^\dagger,	
\end{align}
where $\hat{o}_s = \int s(\Omega)\tilde{o}(\Omega)d\Omega$, $\tilde{o}$ being any frequency-domain annihilation operator. We use $\ket{\beta}$ to represent a coherent state input in mode $\hat{b}_{is}$. Thus, the input SNR is
\begin{align}
	\mathrm{SNR}_i &= \frac{\braket{\hat{b}_{is}^\dagger\hat{b}_{is}}^2}{\braket{\hat{b}_{is}^\dagger\hat{b}_{is}\hat{b}_{is}^\dagger\hat{b}_{is}}-\braket{\hat{b}_{is}^\dagger\hat{b}_{is}}^2}=|\beta|^2.
\end{align}
Similarly, we can find the output SNR as
\begin{align}
	\mathrm{SNR}_o = \frac{G_1^4(0)|\beta|^4}{2G_1^2(0)G_3^2(0)|\beta|^2+G_1^2(0)|\beta|^2+G_3^2(0)}.
\end{align}
When the scattering loss of the cavity is negligible, we have $G_3(0) = \sqrt{G_1^2(0)-1}$. In the high-gain and high-input regime, \textit{i.e.}, $G_1(0), |\beta| \gg 1$, the noise figure is
\begin{align}
	\mathrm{NF} = \frac{\mathrm{SNR}_i}{\mathrm{SNR}_o} \approx 2.
\end{align}
In a critically coupled cavity, we have $G_3(0) = \sqrt{2G_1(0)(G_1(0)+1)}$. In the high-gain and high-input regime, we have 
\begin{align}
	\mathrm{NF} = \frac{\mathrm{SNR}_i}{\mathrm{SNR}_o} \approx 4.
\end{align}

\section{Phase Noise Analysis System.} 
\begin{figure}
	\centering
	\includegraphics{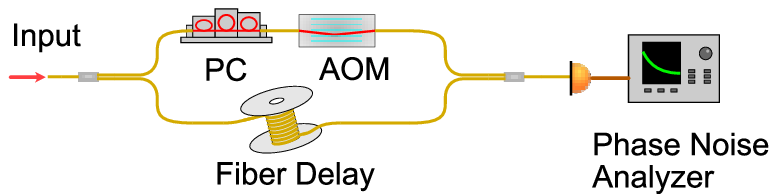}
	\caption{Phase noise analysis system. PC is polarization controller, and AOM is acousto-optic modulator.}
	\label{figPNA}
\end{figure}

We measure the phase noise of lasers using a coherent delayed self-heterodyne system \cite{Camatel_JLT_2008}. As shown in Fig. \ref{figPNA}, the system consists of an arm-length imbalanced Mach-Zehnder interferometer. The fiber delay is 100 m which is shorter than the coherence length of the lasers we measure. We drive the acousto-optic modulator (AOM) at 80 MHz, which has a much lower phase noise than the lasers. Thus, the phase noise of the resulting beat signal is 
\begin{align}
	S_\mathrm{rf}(\omega) = 4\sin^2(\frac{\omega\tau}{2})S_\mathrm{opt}(\omega),
\end{align}
where $\tau$ is the time delay corresponding to the arm-length difference.

\section{Peak-Gain Tuning with Coupled Cavities}\label{sec_Tuning}
\begin{figure}
	\centering
	\includegraphics{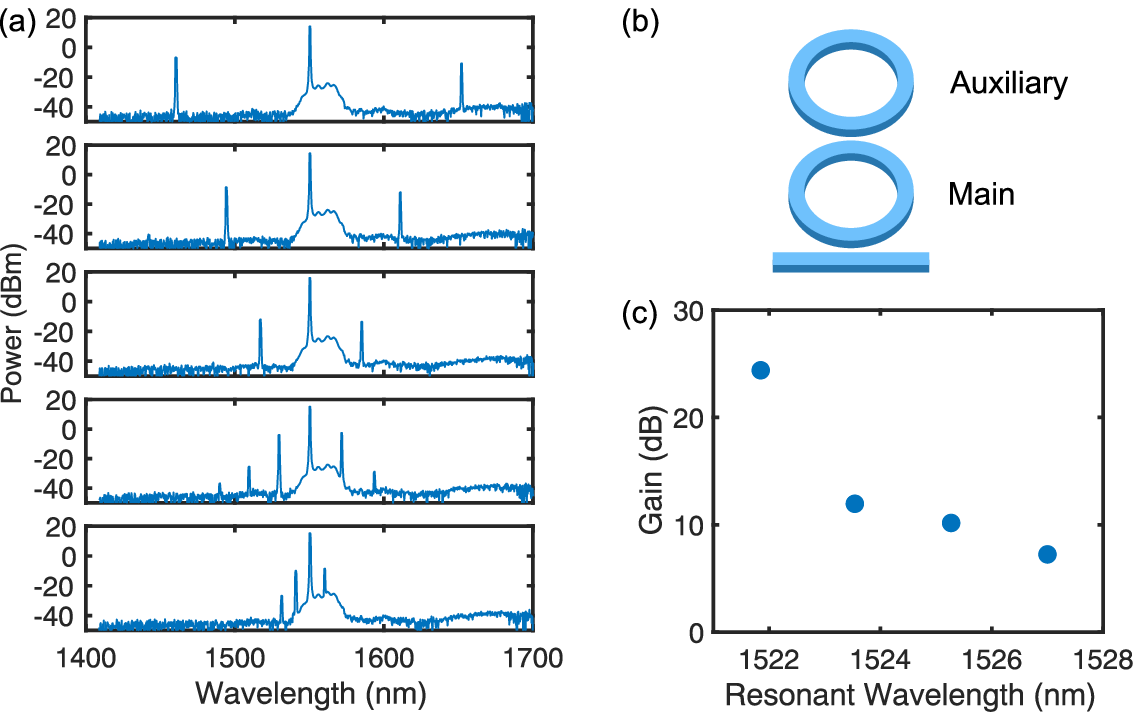}
	\caption{Peak gain tuning and gain at non-peak resonances. a, The change of peak-gain location by coupled-ring tuning. b, The coupled-ring structure. c, Comparison of gain at peak and non-peak resonances.}
	\label{figDualRing}
\end{figure}
Resonances from two cavities with similar frequencies strongly interact when an optical coupling is introduced between them, leading to shifts in their resonance frequencies. For cavities with different free-spectral ranges (FSRs), such strong mode interactions occur locally, allowing phase matching of optical-parametric-oscillation (OPO) processes at the mode interaction region. Since the strength of the localized resonance shift depends on the separation of resonance frequencies, they can be actively tuned via on-chip heaters, which is equivalent to tuning the peak optical-parametric-amplifier (OPA) gain location. We experimentally show this using coupled microresonators with a device similar to that in \cite{Kim_OL_2019}. We use a cross-section of 640 $\times$ 1800 nm, a main-ring FSR of 200 GHz, and an auxiliary-ring FSR of 201.6 GHz (Fig. \ref{figDualRing}b). The pump power used in the experiment is 100 mW, which is due to the effective loss introduced by mode coupling. Such loss can be mitigated by adjusting the coupling strength and frequency difference of the rings \cite{Zhao_OL_2021}. We generate OPOs at disparate wavelengths by tuning the auxiliary-ring heater [Fig. \ref{figDualRing}(a)] while keeping the main-ring heater at a constant voltage. The peak gain on the short-wavelength side shifts from 1460.6 nm to 1540.7 nm, and the peak gain on the long-wavelength side shifts from 1652 nm to 1560.5 nm, corresponding to a  10.7 THz tuning on each side of the pump.

\section{Gain Coefficient at Different Resonances}
We characterize the OPA gain for different resonances under the same pumping condition. We use a pump power of 9 mW and a seed power of 650 nW. The peak gain is at 1521.8 nm, which corresponds to the OPO wavelength. Due to the limited tuning range of our seed laser, we probe the gain for wavelengths larger than the peak gain wavelength. Figure \ref{figDualRing}(c) shows the different gain factors for different cavity resonances. We see that the gain is significantly lower for resonances other than the peak-gain resonance. This is because these resonances are farther away from oscillation. Nonetheless, we observe a 7-dB gain for the resonance 5-nm away from the peak-gain location.

\section{Pump and OPO Noise}
\begin{figure}
	\centering
	\includegraphics[scale=0.5]{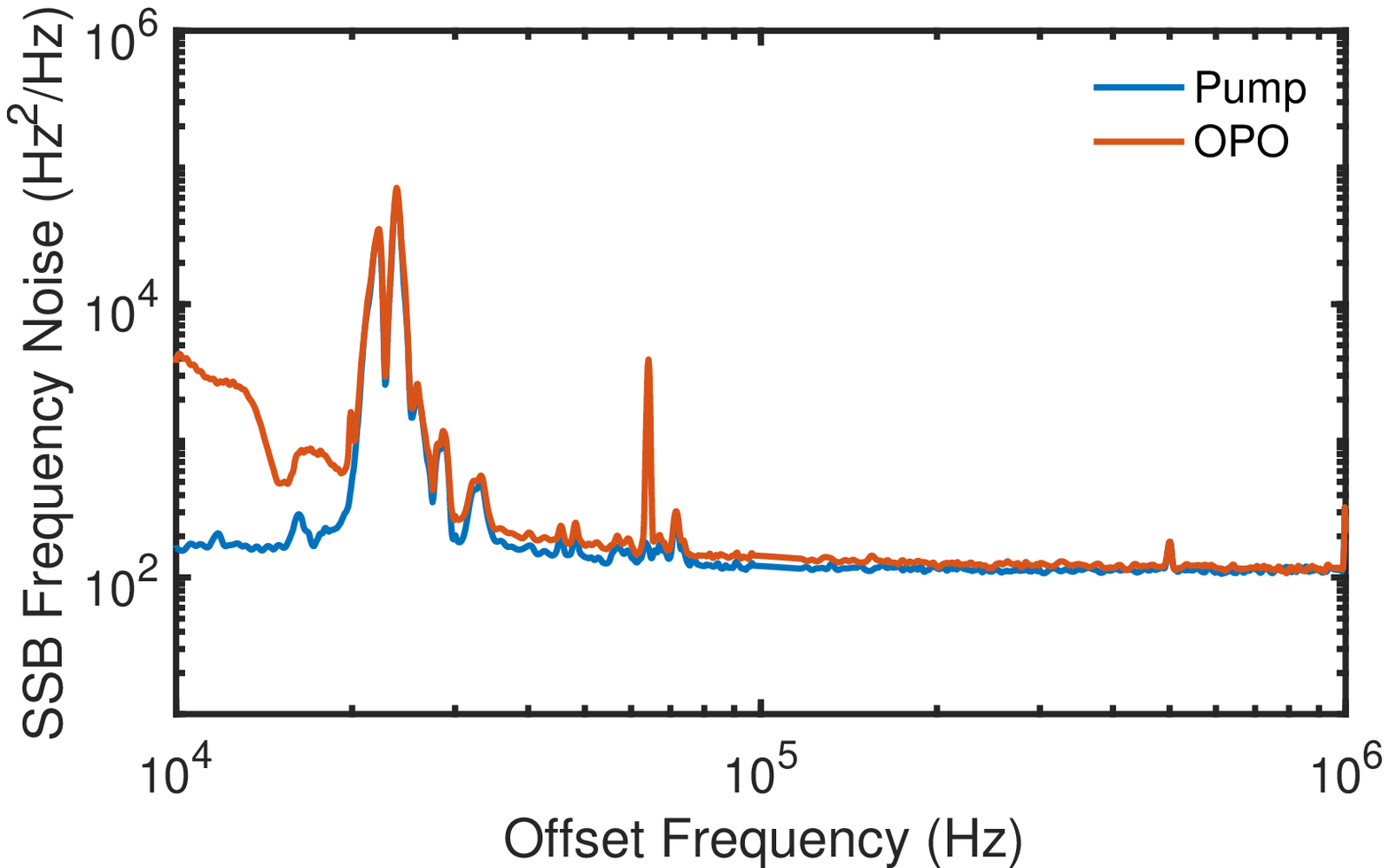}
	\caption{Comparison of pump and OPO frequency noise. The OPO noise is dominated by thermal noise at low offset frequency, and pump noise at high frequencies and around 23 kHz where pump noise is high.}
	\label{figPumpNoise}
\end{figure}
The phase noise analysis system shown in Methods is used to characterize the frequency noise of the pump and OPO. As shown in Fig. \ref{figPumpNoise}, The OPO noise agrees with pump noise at frequencies above 100 kHz where thermorefractive noise is low. Additionally, the OPO noise is dominated by pump noise around 23 kHz where the pump noise has a strong peak. At lower offset frequencies, the thermorefractive noise dominates the OPO noise. Finally, the spurious noise peak of the OPO at 65 kHz is due to the voltage supplier connected to the integrated heater.

\section{Application of Low-Pump-Power Narrow-Bandwidth OPAs}
\begin{figure}
	\centering
	\includegraphics{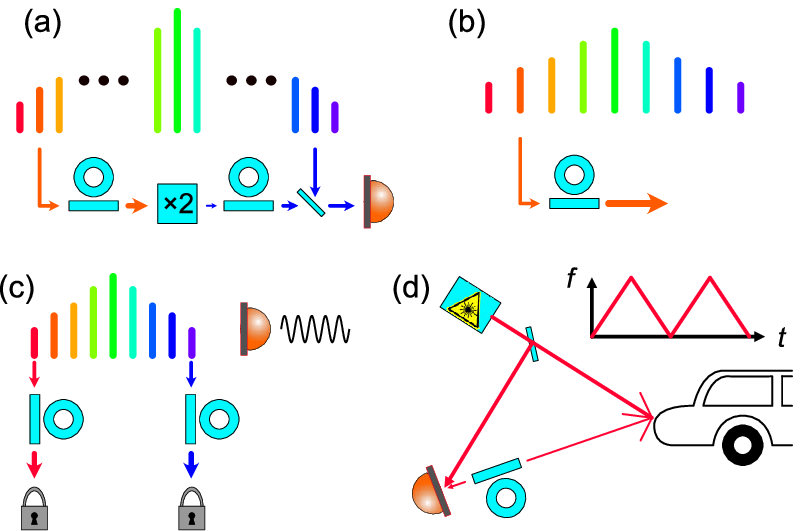}
	\caption{Applications of microresonator-assisted parametric amplification in time and frequency metrology. The OPAs are represented by ring resonators. a, The $f$-$2f$ technique for stabilizing a Kerr frequency comb. A low-frequency comb line is amplified and frequency doubled to overlap with a high-frequency comb line. The frequency-doubled signal requires additional amplification due to the low-efficiency doubling. b, An optical-frequency synthesizer. The Kerr comb has stabilized CEO and repetition frequencies, allowing precise knowledge each comb line's frequency. The OPA is used to increase the power of the desired frequency output. c, Microwave generation using Kerr frequency comb. Two comb lines of a Kerr comb are locked to two stable frequency references, resulting in an ultrastable repetition frequency in the microwave domain. The two comb lines need to have large separations for sufficient noise reduction. The quality of comb-line locking depends on its power which can be significantly improved with microresonator-assisted OPA. d, Illustration of FMCW LIDAR. The probe laser is frequency modulated (inset), reflected from a distant object, and a small portion is collected by the receiver. The received light is mixed with a portion probe laser at the receiving time, which have slightly different frequencies. The time delay can be found by detection the beat note of the light. Due to the typical low-level of the received light, an OPA can be used to improve the signal level, which consequently reduces the required probe power.}
	\label{figSupApplications}
\end{figure}
To achieve parametric amplification at a power level compatible with typical chip-based lasers, the bandwidth of our OPA and injection-locking schemes are generally limited to a sub-gigahertz level. However, the center of the gain is highly adjustable either passively by designing the dispersion before fabrication, or actively using the coupled ring structure mentioned in the first section. The ability to provide gain in a wide wavelength range based on the same material platform makes our scheme highly compatible with frequency-comb-related time and frequency metrology applications. In this section, we briefly review some tasks in on-chip time and frequency metrology and show how our OPA can be applied to facilitate the eventual integration of the devices.

The frequency of the n\textsuperscript{th} comb line of a frequency comb can be expressed as,
\begin{align}
	f_n	= f_\mathrm{CEO}+nf_\mathrm{rep},
\end{align}
where $f_\mathrm{CEO}$ is the carrier-envelope-offset (CEO) frequency, and $f_\mathrm{rep}$ is the repetition frequency. A fully stabilized comb requires knowledge of the CEO frequency, which is usually in the microwave domain. The CEO of a comb can be measured using the $f$-$2f$ interferometry, which requires an octave-spanning comb. Such a comb can be generated on-chip via Kerr-soliton formation \cite{Pfeiffer_Optica_2017,Yu_PRAppl_2019}. As shown in Fig. \ref{figSupApplications}(a), a comb line on the low-frequency side is frequency doubled and mixed with a comb line on the high-frequency side for detection, where the beat note corresponds to the CEO frequency. Due to the nature of soliton modelocking, the comb power at the two wings is generally in the sub-microwatt level, which requires amplification before frequency doubling. For an octave-spanning comb pumped by a telecom-band laser at 1.55 $\mu$m, the two wings correspond to 2.32 $\mu$m and 1.16 $\mu$m, which can be challenging to find gain materials. Our OPA scheme provides a promising solution to such comb-line-amplification tasks. Furthermore, due to the lack of intrinsic $\chi^{(2)}$ nonlinearity, frequency doubling on the SiN platform requires either heterogeneous integration of $\chi^{(2)}$ materials \cite{Rabiei_OE_2013} or the photogalvanic effect \cite{Anderson_OL_1991}, which can lead to additional insertion losses or low conversion efficiencies. However, owing to the low-noise nature of our microresonator-assisted OPA, it can be used to amplify the high-frequency signal after inefficient doubling with minimum added noise, which can enable a high signal-to-noise ratio in beat note detection. The detected beat note can be used to stabilize the CEO frequency with a feedback mechanism. A fully stabilized comb also requires the stabilization of repetition frequency, which can be achieved by directly measuring the pulse train with a photodetctor or stabilizing an optical frequency. The different approaches correspond to different applications.

One application of a fully stabilized comb is optical-frequency synthesis, which outputs a precise optical tone at the requested frequency. In such an application, the CEO frequency is stabilized using the $f$-$2f$ interferometry, and the repetition frequency is stabilized via pulse-train detection. Such a stabilzation process uniquely determines the exact frequency of each comb line. With the broad tunability of our OPA, it can be tuned into the desired frequency to boost the output power of the frequency synthesizer, which otherwise is at microwatt output-power levels. Multiple OPA rings that share the same pump laser can be fabricated to cover an even larger frequency range. 

The repetition frequency of the comb can also be stabilized by locking an optical frequency to an atomic transition, which is the basis of optical atomic clocks \cite{Ludlow_RMP_2015}. In such a case, the repetition frequency is determined to an accuracy of ,
\begin{align}
	\Delta f_\mathrm{rep} = \frac{\Delta f_n-\Delta f_\mathrm{CEO}}{n},
\end{align}
where $\Delta f_\mathrm{rep}$, $\Delta f_n$, and $\Delta f_\mathrm{CEO}$ are the uncertainties of the repetition, optical, and CEO frequencies, respectively. for a 10-GHz repetition frequency, the division factor $n$ is $\approx$ 20000, which results in an extremely precise microwave frequency standard. Besides the OPAs that can be used for CEO stabilization, on-chip optical atomic clock can also benefit from an OPA by amplifying the comb line used to probe the atomic resonance. 

High-frequency and low-noise microwave oscillator is also a highly sort after technology, with applications ranging from telecommunication to sensing. Instead of using a CEO-stabilized comb which requires octave-spanning and a frequency-doubler, microwave generation can be achieved using a dual-point reference scheme. As shown in Fig. \ref{figSupApplications}(c), two comb lines from a Kerr comb are filtered out and locked to narrow-linewidth references, which can be lasers \cite{Li_Science_2014,Tetsumoto_NatPhot_2021}or compact cavities \cite{McLemore_arXiv_2022}. Similar to optical atomic clocks, the two references should have large separations to achieve a large noise suppression factor, which results in low comb-line powers. The comb lines can be boosted by the OPA to enable a high locking quality.

Finally, our OPA scheme can also find important applications in frequency-modulated-continuous-wave (FMCW) LIDAR systems. As shown in Fig. \ref{figSupApplications}(d), a cw laser with frequency modulation (inset) is sent out as the probing signal. The probe laser is scattered upon incident on an object, and a small portion of the scattered light is collected by the receiver. The received light is mixed with a local copy of the probe laser (local oscillator), which has a small frequency difference due to the delay between the emission and receiving time. The frequency difference can be resolved by performing a Fourier transform of the temporal trace, which can be used to infer the distance between the emitter and the object. Due to the high accuracy of frequency measurement, the FMCW LIDAR has a much higher distance resolution than the time-of-flight LIDAR. However, a relatively high probe power is needed due to the low efficiency of collecting scattered light. Our OPA system can provide a large amplification to the received weak light without adding appreciable frequency noise. This can be achieved by tuning the OPA ring together with the probe laser so that the amplification band does not strongly deviate from the probe frequency, which is sufficient to cover the typical frequencies of the received light. Our OPA scheme can allow FMCW LIDAR to operate at a much lower probe power level. 

\section{Simulation for Ultraviolet and Midinfrared OPA}
\begin{figure}
	\centering
	\includegraphics{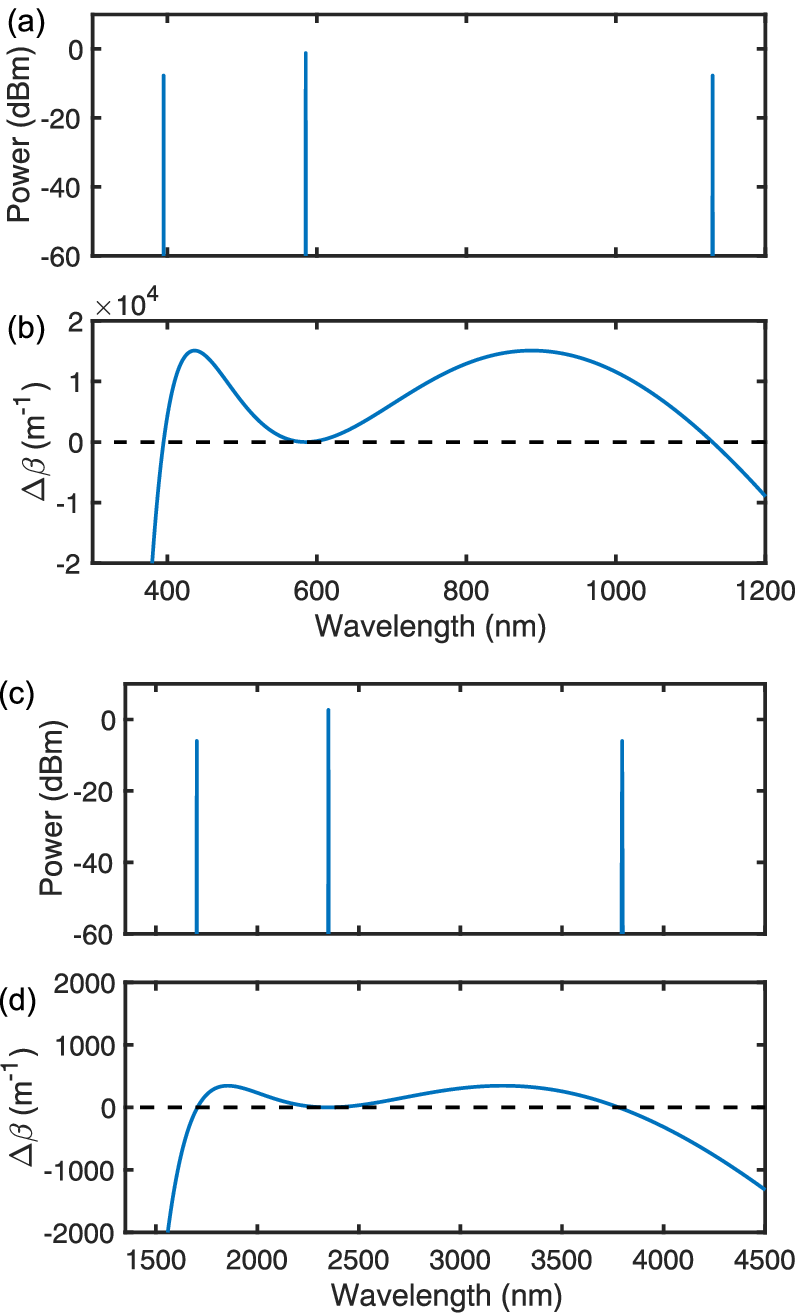}
	\caption{Simulation of parametric gain in the ultraviolet and midinfrared regimes. The parametric oscillation (a) and phase matching (b) of the TM$_{01}$ mode in an air-cladded silicon-nitride microresonator with a cross-section of 730 $\times$ 1100 nm$^2$. The parametric oscillation (c) and phase matching (d) of the TE$_{00}$ mode in a silica-cladded silicon microresonator with a cross section of 730 $\times$ 2000 nm$^2$.}
	\label{figSupUVMidIR}
\end{figure}
Waveguide dispersion engineering greatly expands the spectral coverage of OPAs. In particular, phase matching via higher-order dispersion allows parametric gain far from the pump wavelength. In this section, we show numerical simulations of parametric gain at the ultraviolet and midinfrared regimes pumped by yellow and telecom-U-band lasers respectively.

The generation of parametric gain requires the balanced of dispersive and nonlinear phases. In the case of gain near the pump wavelength, the dominant dispersive effect is the group-velocity dispersion (GVD) which must be anomalous for a degenerately pumped four-wave-mixing (DFWM) process. However, if gain far from the pump is desired, higher-order dispersive effects need to be considered \cite{Harvey_OL_2003,Lu_NatPhot_2019,Sayson_NatPhot_2019,Domeneguetti_Optica_2021}. Due to the symmetric locations of the signal and idler, only even order dispersive effects contribute to the phase matching of DFWM. In general, the phase-matched frequencies satisfy \cite{Sayson_NatPhot_2019},
\begin{align}
	\Delta\beta = \beta_2\Delta\omega^2+\frac{\beta_4}{12}\Delta\omega^4 \approx 0,\label{eq_beta4}
\end{align}
where $\Delta\beta$ is the phase mismatch, $\beta_2$ is the GVD coefficient, $\beta_4$ is the 4\textsuperscript{th}-order dispersion, and $\Delta\omega$ is the frequency separation from the pump. We use the mean-field Lugiato-Lefever equation \cite{Coen_OL_2013} to show that parametric oscillation can be excited in the ultraviolet and midinfrared regimes. Since OPO corresponds to parametric gain exceeding the roundtrip loss, the OPA condition where the parametric gain exceeds the intrinsic loss can also be satisfied.

Due to the strong normal GVD of photonic materials at short wavelengths, the phase matching of DFWM requires the assistance of equally strong waveguide dispersion, which can be achieved via high-order waveguide modes \cite{Zhao_Optica_2020,Domeneguetti_Optica_2021}. We simulate an air-cladded silicon-nitride microresonator with a cross-section of 730 $\times$ 1100 nm$^2$ and a radius of 100 $\mu$m. We choose to operate in the TM$_{01}$ mode which corresponds to a $\beta_2 = 2.5\times10^{-26}$ s$^2$/m and $\beta_4 = -1.3\times10^{-55}$ s$^4$/m. We assume critically coupled resonances with a linewidth of 200 MHz, a pump wavelength of 585 nm, and power of 30 mW. Equation (\ref{eq_beta4}) predicts a phase-matching wavelength of 395 nm [Fig. \ref{figSupUVMidIR}(b)] which agrees with the OPO wavelength.

For parametric gain in the midinfrared regime, we use the fundamental TE mode of a silica-glass-cladded silicon waveguide similar to \cite{Griffith_NatComm_2015}, with a cross section of 500 $\times$ 2000 nm$^2$ and a radius of 100 $\mu$m which corresponds to $\beta_2 = 1.5 \times 10^{-26}$ s$^2$/m and $\beta_4 = -2.0 \times 10^{-54}$ s$^4$/m. We also assume a critically-coupling condition with a linewidth of 400 MHz, a pump wavelength of 2350 nm, and power of 22 mW. We find the peak gain at 1701 nm and 3796 nm in this system [Fig. \ref{figSupUVMidIR}(d)].

\end{bibunit}

\end{document}